%% file: main.tex
\documentclass[submission,copyright,creativecommons]{eptcs}
\providecommand{\event}{FMAS 2024} 

\usepackage{iftex}
\usepackage{graphicx}

\ifpdf
  \usepackage{underscore}         
  \usepackage[T1]{fontenc}        
\else
  \usepackage{breakurl}           
\fi

\title{Open Challenges in the Formal Verification\\ of Autonomous Driving}
\author{Paolo Burgio \quad \quad Angelo Ferrando \quad \quad Marco Villani
\institute{University of Modena and Reggio Emilia\\ Department of Physics, Informatics and Mathematics\\Modena, Italy}
\email{forename.surname@unimore.it}
}
\def\titlerunning{Open Challenges in the Formal Verification of Autonomous Driving}
\def\authorrunning{P. Burgio, A. Ferrando \& M. Villani}
\begin{document}
\maketitle

\begin{abstract}
In the realm of autonomous driving, the development and integration of highly complex and heterogeneous systems are standard practice. Modern vehicles are not monolithic systems; instead, they are composed of diverse hardware components, each running its own software systems. An autonomous vehicle comprises numerous independent components, often developed by different and potentially competing companies. This diversity poses significant challenges for the certification process, as it necessitates certifying components that may not disclose their internal behaviour (black-boxes). In this paper, we present a real-world case study of an autonomous driving system, identify key open challenges associated with its development and integration, and explore how formal verification techniques can address these challenges to ensure system reliability and safety.
\end{abstract}

\section{Introduction}

The Society of Automotive Engineers (SAE) defines the design goals of autonomous driving across six distinct levels, ranging from Level 0 (L-0) to Level 5 (L-5), as outlined in~\cite{sae2021taxonomy}. These levels represent a spectrum of automation: L-0 denotes no automation, followed by L-1 which includes driver assistance, L-2 for partial automation, L-3 for conditional automation, L-4 for high automation, and culminating in L-5, which signifies full automation. Each level reflects the increasing capability of autonomous systems and their interaction with human drivers.
Currently, most commercially available vehicles operate at L-2 automation. This level encompasses features such as adaptive cruise control and lane-keeping assistance, enabling the vehicle to assist the driver while still requiring constant supervision and active engagement. A few manufacturers are exploring L-3 systems, which offer conditional automation under specific circumstances. For example, some modern vehicles equipped with L-3 systems can handle highway driving autonomously, including lane-keeping, speed regulation, and adaptive cruise control, but require the driver to take over when exiting highways or in complex urban environments. Despite such advancements, the industry is still in the early stages of fully implementing higher levels of automation.

According to a recent survey on the subject~\cite{DBLP:journals/csur/KhanEMZKAI23}, achieving L-5 autonomy requires the appropriate integration of technologies and efficient communication channels. Realising the full potential of automated driving demands a \textbf{reliable}, robust, and widespread mobile network. In this work, we focus on the first item on the list; that is, we are interested in making the components of autonomous driving, as well as their interactions, (more) reliable. To achieve this, we start with a case study of a real-world autonomous driving system and address the issues to enhance its reliability from a formal perspective.

Taking inspiration from~\cite{DBLP:journals/corr/abs-2208-05507}, we treat the autonomous driving system as a component-based system composed of black-box components that we are not interested in opening (or cannot open). Instead, we focus on how to achieve the formal verification of the resulting heterogeneous system. That is, how the components interact with each other, and how we can verify (and perhaps even enforce) correct behaviour according to well-known standards in autonomous driving.

This paper presents a real-world case study in autonomous driving (Section~\ref{sec:casestudy}), highlighting key challenges. Section~\ref{sec:fm} examines how formal methods, particularly formal verification, can address these issues and be integrated into autonomous systems, with Section~\ref{sec:fm-limits} discussing their limitations. Finally, Section~\ref{sec:conclusions} concludes the paper and outlines future directions.

\input{casestudy}

\section{Formal Methods to the Rescue}
\label{sec:fm}


In the previous section, we discussed a real-world case study in autonomous driving, highlighting challenges in integrating autonomous systems into road infrastructure and progressing towards L-5 capabilities. Here, we explore how formal methods, particularly formal verification, address these challenges. For a comprehensive overview of formal verification techniques in autonomous systems, see~\cite{DBLP:journals/csur/LuckcuckFDDF19}.

\subsection{Open Challenge 1: Heterogeneous Composition of Untrustworthy Components}

The first challenge involves managing components with varying levels of reliability. Some components are open-source (white-box), allowing full access and modifications, while others are closed-source (black-box), restricting access to their internal workings. To address this challenge, we propose exploiting formal verification techniques. Specifically, as outlined in~\cite{doi:10.1177/1748006X211034970}, we employ formal verification methods focusing on three key areas (called also recipes): verification of decision-making components, AI-based components, and the enforcement of safety claims.

To address this challenge, we propose the use of heterogeneous verification techniques~\cite{DBLP:journals/corr/abs-2208-05507,DBLP:journals/fteda/BenvenisteCNPRR18,DBLP:conf/sefm/ChampionGKT16,DBLP:conf/fm/RuchkinSISG18}.
These techniques are based on the Assume-Guarantee principle, where each component of the heterogeneous system is defined in terms of its \textit{assumptions} (what the component expects from the system to function correctly) and its \textit{guarantees} (what the component provides to the system upon correct execution). This methodology allows us to abstract away the implementation details of the various system components, enabling the system designer to focus on their integration. As long as the assumptions and guarantees of a component are documented and made available, it can be implemented as either a white-box or black-box component. By employing these verification techniques, it is possible to formally verify the proper integration of multiple components, potentially developed by different parties~\cite{DBLP:journals/corr/abs-2208-05507,DBLP:journals/fteda/BenvenisteCNPRR18,DBLP:conf/sefm/ChampionGKT16,DBLP:conf/fm/RuchkinSISG18}.

In addition to Assume-Guarantee reasoning, model checking and formal methods can play a crucial role in verifying component integration and ensuring system reliability~\cite{Clarke1999,DBLP:journals/iee/BensalemBNS10,DBLP:journals/sosym/FalconeJNBB15,DBLP:journals/dafes/KarlssonEP07}. Component-Based Software Engineering (CBSE) methodologies also provide a framework for assembling reliable systems from diverse components~\cite{DBLP:books/lib/SzyperskiGM02}. Moreover, adhering to safety and certification standards, such as ISO 26262, is essential for validating the safety of automotive software systems~\cite{ISO26262}.

To complete the verification process, we envision the use of Runtime Verification (RV)~\cite{DBLP:series/lncs/BartocciFFR18}, a technique for monitoring and analysing the execution of a system at runtime to ensure it adheres to specified properties. RV can check and enforce adherence to all assumptions and guarantees of the components. As highlighted in~\cite{DBLP:journals/corr/abs-2208-05507}, the Assume-Guarantee verification methodology focuses on verifying the resulting distributed system and the integration of its components. However, it relies on the assumption that the assumptions and guarantees of each component (which may be black-boxes) are satisfied. To bridge this gap and provide a robust verification technique suitable for the heterogeneous nature of the autonomous driving domain, we need to employ additional verification methods to ensure the proper behaviour of individual components. By confirming that each component behaves correctly, we can validate the entire system's integration and maintain its formal assurances.

It is important to note that the use of RV in this context is not entirely straightforward. The components of an autonomous driving system may exhibit a certain level of uncertainty, meaning that the information they provide may not always be precise. For example, as illustrated in Figure~\ref{fig:citybox-scheme}, the steps that process the camera input to determine the driver's level of drowsiness, distraction, fatigue, etc., are inherently uncertain. These steps rely on Machine Learning models, which offer results with varying levels of confidence. Additionally, this uncertainty is not limited to the current information provided but may also encompass temporal aspects. For instance, a component might predict that the driver will fall asleep in five minutes, with a given level of confidence.
Due to these factors, it is unrealistic to rely solely on standard RV approaches for verifying component conformance. Instead, techniques that incorporate RV with uncertainty must be considered, such as those discussed in~\cite{DBLP:conf/rv/WangASL11,DBLP:conf/icse/TalebKH21,DBLP:conf/sefm/FerrandoM22}. For a comprehensive survey on this topic, the reader may refer to~\cite{DBLP:journals/csr/TalebHK23}.
Additionally, the complexity of verifying machine learning and AI components within autonomous systems presents unique challenges. Ensuring the reliability of non-deterministic algorithms requires specialised verification techniques~\cite{Marcus2019}. Addressing these challenges will enable the development of robust, reliable, and safe autonomous driving systems.

\subsection{Open Challenge 2: Efficient Strategies for Mapping the Distributed Computation}

Addressing this challenge involves the use of formal verification techniques to ensure that the mapping strategies are both efficient and reliable (since we are in a real-time system). Formal verification can be employed to systematically verify that the software components are optimally distributed across the computing cores, and that redundancy and voting mechanisms are correctly implemented to enhance fault tolerance and system robustness. By formally verifying these strategies, we can guarantee the system meets its performance and reliability requirements, even in the presence of component failures or uncertainties.
Indeed, fault-tolerant designs, which incorporate redundancy and voting schemes, play a crucial role in mitigating the impact of component failures. The study presented in~\cite{DBLP:journals/access/EntrenaSGPGLS23} provides valuable insights into the application of formal verification techniques to validate the correctness of these designs. By systematically verifying that fault-tolerant hardware meets specified reliability requirements, the authors demonstrate the effectiveness of formal methods in identifying design errors that traditional testing might overlook.
Although~\cite{DBLP:journals/access/EntrenaSGPGLS23} does not originate from the domain of autonomous driving, it provides a valuable foundation for addressing the open challenge discussed here. The paper presents methodologies for formal verification of fault-tolerant hardware designs, which are crucial for ensuring the reliability and robustness of systems with heterogeneous components. By adapting these verification techniques, we can systematically validate the correctness and reliability of the complex, integrated systems used in autonomous vehicles. 

Formal verification can be used to prove that the redundancy and voting mechanisms are correctly implemented and that they effectively enhance system reliability; for example, in~\cite{DBLP:journals/access/SafariAKGYYHEH22}, the authors discuss fault-tolerance techniques that include redundancy and efficient scheduling policies. Formal verification ensures that these techniques are correctly applied, thereby enhancing the reliability of the system.

The work in~\cite{DBLP:conf/date/IqtedarHSH16} provides a comprehensive framework for the formal verification of distributed Resource Management (RM) schemes in many-core systems using probabilistic model checking. This research is particularly relevant to our work in the context of autonomous driving systems, which also require efficient resource allocation across multiple computing cores. The authors demonstrate the use of the PRISM model checker~\cite{DBLP:conf/cpe/KwiatkowskaNP02} to analyse and compare the performance and reliability of different RM schemes. They emphasise the limitations of traditional simulation methods, which are inherently exhaustive, and advocate for formal verification to ensure completeness and accuracy. In our study, similar formal verification techniques can be applied to optimise the mapping of software components onto the available computing cores in autonomous vehicles. By leveraging the probabilistic analysis methods described in~\cite{DBLP:conf/cpe/KwiatkowskaNP02}, we could systematically evaluate the robustness and performance efficiency of our proposed resource management strategies in autonomous driving systems.

\section{Limitations of Applying Formal Methods in Autonomous Systems}
\label{sec:fm-limits}

While Formal Methods provide a promising approach for integrating heterogeneous components and efficient mapping in autonomous systems, there are significant practical limitations to their use. The main challenges include the need for specialised knowledge, scalability issues, interpretability of results, and difficulties in handling uncertain environments, as well as cost-benefit trade-offs in development.

One key limitation is the high barrier to entry, as FM often requires deep expertise in formal logic and verification techniques. This specialised skill set is not commonly available within standard engineering teams, and the process of specifying and verifying systems can be time-consuming. Despite the development of new tools aimed at making FM more accessible, their capabilities are still evolving and often require substantial refinement to meet industry needs.

Scalability and complexity also present major obstacles. Autonomous systems comprise numerous interacting components, leading to a state space that grows exponentially, making exhaustive verification computationally expensive or infeasible. Techniques like compositional reasoning and modular verification attempt to manage this, but they require careful abstraction, which may oversimplify or overlook critical behaviours. Moreover, model checking can be computationally intensive, particularly when applied to real-time or resource-constrained systems, and Assume-Guarantee reasoning hinges on accurate assumptions that are challenging to guarantee in practice.

Another practical challenge is the interpretability of verification results. Outputs from FM tools, such as model checkers or runtime verifiers, may highlight specification violations without providing clear solutions, requiring domain expertise to resolve. When AI and machine learning components are involved, this issue is further complicated by their probabilistic and non-deterministic nature, making the analysis of verification results particularly difficult.

The dynamic and uncertain environment in which autonomous systems operate adds further complexity. Perception modules relying on sensor fusion and AI-based decision-making are context-dependent and produce inherently uncertain outputs. Traditional FM approaches struggle to define precise specifications in such cases. Techniques that incorporate probabilistic reasoning or uncertainty-aware models are being explored~\cite{DBLP:conf/rv/WangASL11,DBLP:conf/icse/TalebKH21,DBLP:conf/sefm/FerrandoM22}, but their practical applicability to real-world systems remains under active research and development.

Integrating FM into existing development workflows also poses a significant challenge, as it requires a careful cost-benefit analysis. The process of formally specifying, modeling, and verifying components demands significant time and resources, potentially extending the development lifecycle. While FM provides strong safety and reliability assurances -- critical for autonomous vehicles -- these benefits must be weighed against their scalability and adaptability to system updates and changes.

Lastly, verifying AI and ML components remains an open challenge, as their non-deterministic behaviour does not fit within traditional FM frameworks. Specialised techniques are needed to manage varying levels of confidence and probabilistic outputs in AI models~\cite{Marcus2019}. Thus, hybrid approaches combining formal verification with testing and validation are necessary for comprehensive system assurance.

In summary, FM offers a rigorous foundation for ensuring system reliability and safety in autonomous vehicles, but practical limitations such as scalability, required expertise, result interpretation, and integration into dynamic environments must be addressed to enable their effective use in real-world applications.

\section{Conclusions and Future Work}
\label{sec:conclusions}

In this short paper, we present a real-world case study in the autonomous driving domain, identify key open challenges, and discuss how formal verification techniques can address these issues. 

We focus on two primary challenges hindering the achievement of L-5 automation in autonomous vehicles. The first is the heterogeneous composition of black-box components, which affects system reliability. The second challenge involves the proper mapping of computations within a highly dynamic, real-time distributed system. We propose how existing formal verification techniques can be leveraged to address these issues and explore their implications, along with potential adaptations for integration into autonomous driving systems.


Our aim is to highlight these open challenges in the autonomous driving domain and demonstrate how formal verification techniques can theoretically enhance system reliability. For future work, we intend to build upon the insights reported here by applying some of the discussed techniques and methodologies to our case study, leveraging tools like those proposed in~\cite{DBLP:conf/woa/0001M24,DBLP:journals/corr/abs-2403-02170} to further explore their practical applicability and impact on system robustness.

\subsection*{Acknowledgements}

The authors have received funding from ECSEL JU project AI4CSM (GA N.101007326) and the Chips JU project ShapeFuture (GA N.101139996).

\bibliographystyle{eptcs}
\bibliography{main}
\end{document}

%% file: casestudy.tex

\section{Autonomous Driving Case Study}
\label{sec:casestudy}

\begin{figure}
    \centering
    \includegraphics[width=0.55\linewidth]{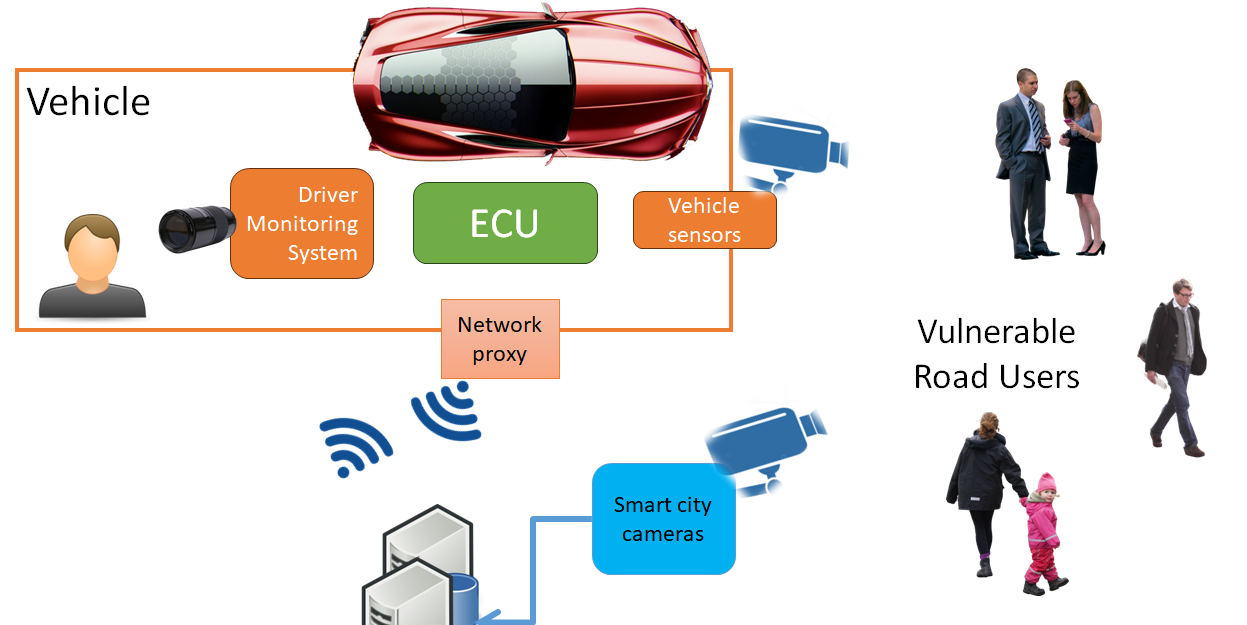}
    \caption{The vehicle architecture and use case}
    \label{fig:usecase}
\end{figure}

As a motivational example, we present a case study from the AI4CSM Project\footnote{https://ai4csm.eu/}, funded by the European Commission~\cite{10.3389/ffutr.2021.688482}. This study features a L-3/L-4 autonomous vehicle equipped with advanced sensor fusion capabilities, exemplifying a next-generation automotive platform~\cite{saesensorfusion, 8943388}. The vehicle can interpret both driver status (e.g., drowsiness, distraction) using in-vehicle cameras, and external environmental conditions using on-board sensors and data from city sensors, as illustrated in Figure~\ref{fig:usecase}.

\begin{figure*}[!ht]
    \centering
    \includegraphics[width=0.55\linewidth]{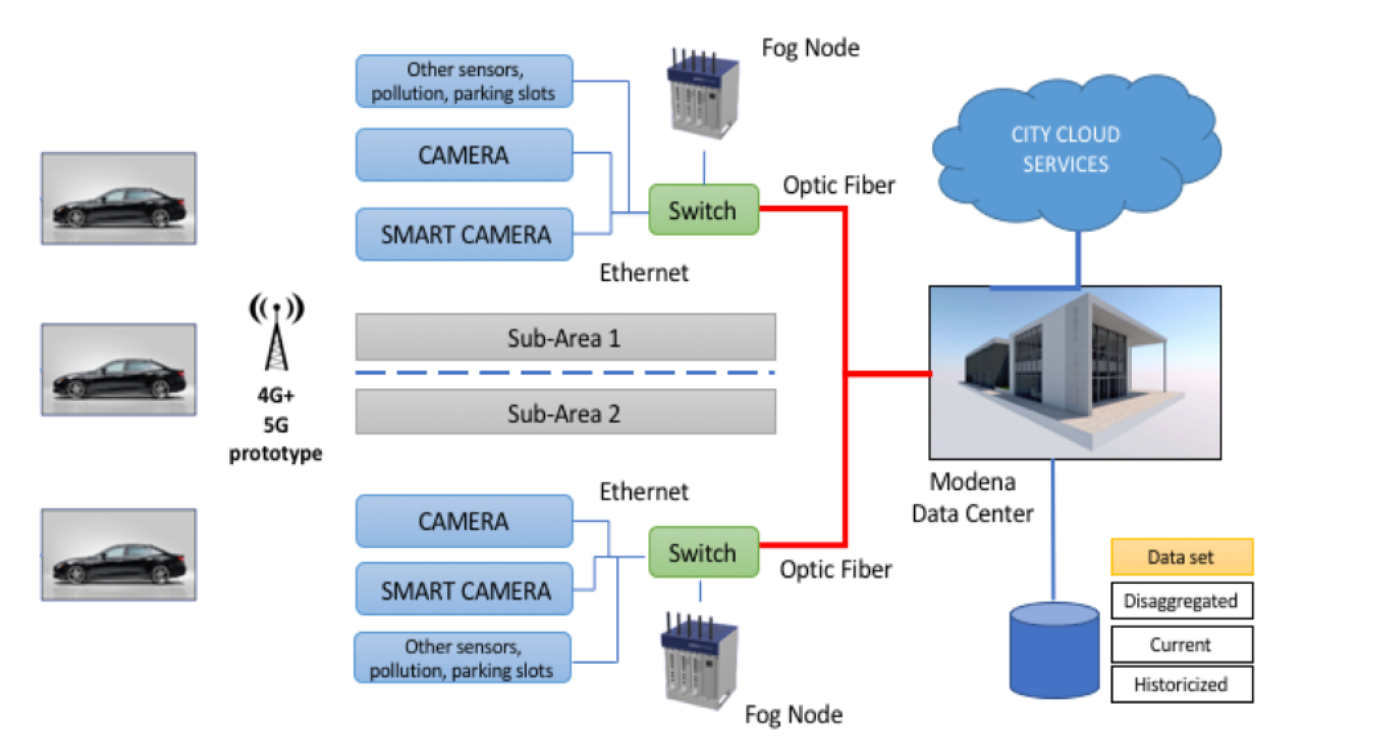}
    \caption{The Modena Automotive Smart Area.}
    \label{fig:masa-arch}
\end{figure*}

In the simplest scenario, the vehicle operates at a low level of automation (\textit{i.e.}, L-2/L-3), with the driver maintaining control. If the vehicle detects a potentially dangerous situation, such as drowsiness or imminent collisions, it triggers a secure takeover strategy, transitioning to L-4 and executing a safety manoeuvre. For our study, we focus on the sensor fusion component, where the perception module running on the on-board Electronic Control Unit (ECU) aggregates information from heterogeneous data streams. Specifically, we implemented a system to monitor the driver using camera-based behavioural analysis, coupled with on-board cameras to inspect the surrounding environment. Additionally, we enhanced the vehicle's perception capabilities by incorporating data from a smart city prototype area, namely the Modena Automotive Smart Area (MASA)\footnote{https://www.automotivesmartarea.it/}. Its structure is shown in Figure~\ref{fig:masa-arch}.

This area includes smart cameras mounted on poles that detect vulnerable road users and analyse or predict their movement trajectories. These data are streamed to the vehicle through the smart city's 4G-5G wireless connectivity. The vehicle’s centralised ECU, also known as the Domain Controller, processes this information to determine the most appropriate response, such as emergency braking, complex manoeuvres, or issuing driver warnings. For research purposes, we implemented this use case on a Citroen Mehari. We will now explore the main open challenges that must be addressed to facilitate the industrialisation of these complex systems.

\begin{figure*}[!ht]
    \centering
    \includegraphics[width=0.7\linewidth]{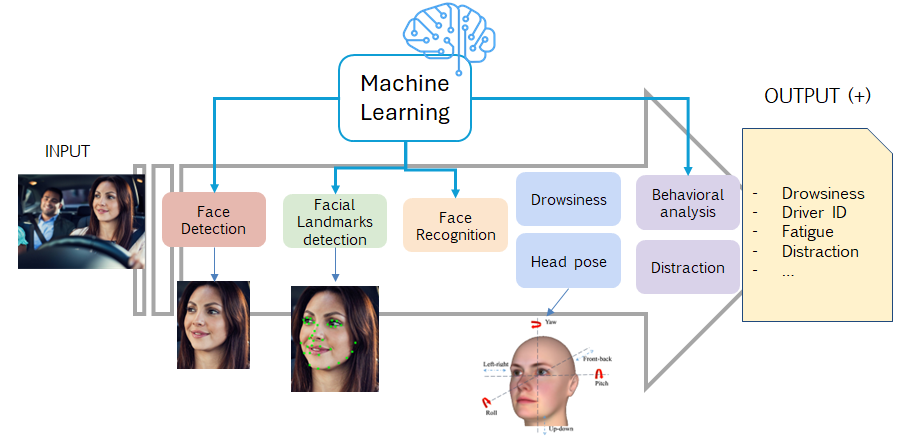}
    \caption{Scheme of the behavioural DMS pipeline.}
    \label{fig:citybox-scheme}
\end{figure*}

\textbf{Aggregating probabilistic data sources}.
One primary open challenge arises from the inherent nature of most software components used for perception. This stage is the first in any autonomous driving stack, where raw sensor data (in our scenario, RGB cameras) are processed to interpret and analyse the driver or the car's surroundings. Numerous algorithms and approaches could be employed, most of which~\cite{saesensorfusion, 8943388, Cavicchioli20223, Liu2018ECCV, Nie2023} heavily rely on machine learning, deep learning, or, more generally, on heuristics and statistical methodologies to handle the complexity of raw data frames. Additionally, most systems have a hierarchical structure, consisting of sub-components arranged in pipelines. Figure~\ref{fig:citybox-scheme} illustrates this decomposition for our camera-based behavioural analysis used for monitoring the driver.

Every block of this system has a specific performance metric, typically expressed in Frames-Per-Second (FPS), and a nominal accuracy, indicating the reliability of the information produced by the (sub)component. Our Driver Monitoring System (DMS) is a white-box component, developed in-house, which allows us complete access to its internals for tuning and modifications. However, in realistic industrial scenarios, most software modules will be developed by different companies and often implemented as ``black-box'' ECUs. Therefore, we identify the first open challenge. 
\begin{center}
\textit{Open Challenge 1: the need to compose a hierarchy of probabilistic software modules to formally measure and derive the overall system's resulting accuracy}.
\end{center}

\begin{figure*}[!ht]
    \centering
    \includegraphics[width=0.4\linewidth]{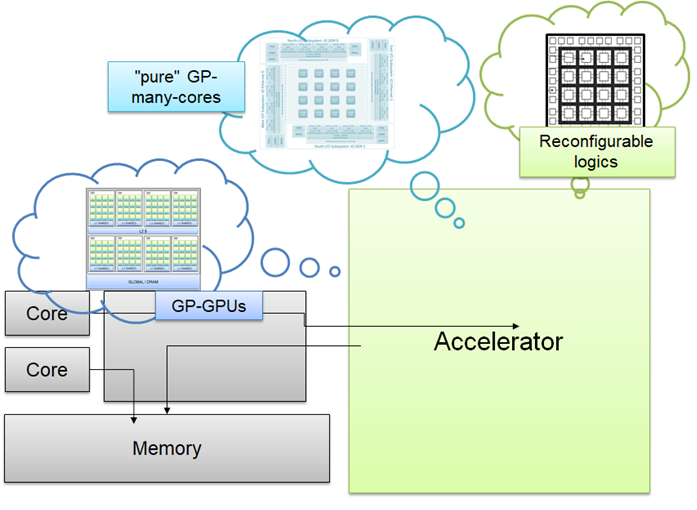}
    \caption{Generic architecture next-generation ECUs.}
    \label{fig:embedded-arch}
\end{figure*}

\textbf{Deploy on embedded systems}.
Deploying intelligence in automation use cases requires two key components: powerful computational hardware and numerous sensor modules to accurately interpret the surrounding and in-cabin environment in a timely manner. This presents a significant challenge for automotive engineers, who must integrate TOPS-greedy\footnote{They require a high number of Tera Operations Per Second (TOPS) to process complex algorithms, such as those used in artificial intelligence and sensor fusion.} software components onto power-efficient boards, ideally featuring many-core data processors such as those from NVIDIA Orin~\cite{orin} or AMD XILINX~\cite{Cho2021AnFE, zcu102}. Figure~\ref{fig:embedded-arch} illustrates the target architecture of next-generation ECUs.

These systems employ multi-core host platforms, which include both Real-Time and non-Real-Time core ISAs (Instruction Set Architectures), which define the set of instructions a processor can execute. These are coupled with data-crunching architectures such as GPGPUs~\cite{orin}, reconfigurable arrays~\cite{zcu102}, or application-specific circuitry to implement processing algorithms directly in hardware. Such a complex architecture presents two main challenges.

\begin{center}
    \textit{Open Challenge 2: to devise efficient strategies for mapping software components onto the available computing cores, exploiting redundancy and voting schemes to enhance overall system reliability.}
\end{center}

Intuitively, most of the algorithms we employ can potentially run on various cores, and finding the optimal mapping must be handled in the most efficient manner.

